\documentclass{elsart}
\usepackage{graphicx}
\usepackage{amssymb}
\usepackage{amsmath}

\begin{document}
\begin{frontmatter}


\title{Local Density of States and Order Parameter Configurations in 
Layered Ferromagnet-Superconductor Structures.
}
\author{Klaus Halterman}
\address{Sensor and Signal Sciences Division, Naval Air Warfare Center,
China Lake, California 93555\thanksref{thanks}}
\ead{klaus.halterman@navy.mil}
\author{Oriol T. Valls}%\corauthref{cor}}
\address{School of Physics and Astronomy and Minnesota Supercomputer
Institute, University of Minnesota, Minneapolis, Minnesota 55455} 
%\corauth[cor]{Corresponding author}
\ead{otvalls@umn.edu}
%\title{}
\thanks[thanks]{This work was supported in part by a grant of HPC resources from 
the Arctic Region Supercomputing Center at the University of Alaska Fairbanks 
as part of the Department of Defense High Performance Computing Modernization 
Program.} 

%\author{}

%\address{}

\begin{abstract}
We analyze the local density of states (LDOS) of heterostructures
consisting of alternating  
ferromagnet, $F$, and superconductor, $S$, layers. 
We consider structures of the
$SFS$ and $SFSFSFS$ type, with thin nanometer scale $F$ and $S$ layers,
within the ballistic regime.
The
spin-splitting effects of the ferromagnet and the
mutual coupling between the $S$ regions,
yield several nontrivial stable and metastable pair amplitude
configurations, and we find that
the details
of the spatial behavior of the pair amplitude govern
the calculated electronic spectra. These
are reflected in discernible signatures
of the LDOS. The roles that
the magnetic exchange energy, interface scattering strength, 
and the Fermi wavevector mismatch
each have on the LDOS 
for the different allowed junction configurations, are 
systematically investigated.

\end{abstract}

\end{frontmatter}
\newpage
% main text
\section{Introduction}
\label{intro}

The continual investigation and development of  heterostructures consisting
of layered ferromagnet ($F$) and superconductor ($S$)
compounds at
nanometer length scales has captured the interest of researchers in 
a broad range of disciplines.
The reasons for this stem  from  
the distinctive properties of these nano-sized materials, 
modified from those of  
their bulk counterparts, related to the effects induced by 
the mutual coupling that occurs when  superconductors  
and ferromagnets are in close electrical contact. 
The advancement of nanofabrication techniques is partly responsible for the
maintained interest thus far, where electron-beam lithography,
etching and lift-off methods, to name a few, 
have been refined to allow for the well controlled creation of relatively clean
multiple layers involving $F/S$  junctions.
Further developments in
experimental probes such as 
the scanning tunneling  microscope (STM),
have also progressed
to the point where $\mu$eV energy resolution can be obtained  at
very low temperatures\cite{courtois}. 
When operating in
spectroscopy mode, the STM gives fine atomic-scale
details of the local density of states (LDOS) of inhomogeneous superconducting systems.
In contrast to conventional tunnel junctions, where transport measurements
are taken, the STM gives
localized spatial information not averaged out over the extent of the entire sample\cite{courtois2}.

For 
composite materials comprised of
multiple superconductor layers separated
by ferromagnet materials, 
the presence of the magnetic $F$ regions
induces a splitting of the Fermi surfaces of the different spin bands,
and  leads to spin-dependent Andreev\cite{and} states. 
The degree of spin-polarization in the superconductor and
the induced pairing correlations in the ferromagnet are
all components that embody what are known as proximity effects.
It turns out that
due to the mutual interaction between the two contrasting 
materials and the inherent system inhomogeneity, the pair amplitude,
becomes 
spatially varying\cite{ff,lo}, and
its phase difference $\Delta \phi$ can modulate between
successive $S$ layers.
For certain configurations, there can be (at zero applied magnetic field) 
a phase difference of
$\Delta \phi=\pi$ between $S$ layers separated by
a ferromagnet\cite{bulaevskii}. These are the so-called $\pi$ junctions,
which occur in the %kh
simplest case, in  a 3-layer $SFS$ type structure.
Junctions of this type can also occur in different combinations %kh
for  more complicated\cite{hv3,izy} layered 
heterostructures,
where the relative sign
of the pair potential $\Delta({\bf r})$ can change between adjacent $S$ layers.
The proximity effect is then
responsible for the existence
of a number of stable junction configurations
of the pair amplitude in $F/S$ multilayers\cite{stab}: the oscillatory
behavior of the superconducting correlations in each ferromagnet and 
the spatial profile of the pair amplitude
in the superconductor are closely interconnected, often resulting in a
nontrivial spatial dependence of 
the magnitude and phase of the order parameter.
This consequently implies that the 
one-particle energy spectrum, or 
density of states (DOS), will have an equally
interesting behavior.

The characteristic behavior of 
the DOS in structures consisting 
of a single ferromagnet and superconductor has been
addressed in past work. In the diffusive, dirty-limit
case, the damped-oscillatory decay of the 
pair amplitude in the $F$ region induces a spatial
variation of the LDOS\cite{buzdin}. For a clean $F/S$ system with 
rough boundaries, the tunneling DOS as a function of energy revealed
certain inversion signatures that depended on the thickness of the
ferromagnet film \cite{zareyan}. 
The energy gap in $F/S$ hybrid systems becomes rapidly suppressed,
in a manner that depends on the ratio of the $F$ and $S$ Fermi energies,
interfacial scattering strength, and $F$ layer thickness\cite{gap1}. 
The depletion of the gap has also been discussed in the context of
a potential spin-valve device \cite{gap2}.
The spatially averaged DOS was calculated for a clean 2-D $F/S$ box 
geometry\cite{koltai}, as
was the LDOS for 3-D hybrid $F/S$ nanostructures\cite{vecino}, where there was a 
reported  filling in of the gap with increased $F$ layer
thickness. The damped oscillatory phenomena of the DOS is expected 
to persist over a broad range of material parameters \cite{buzdin2}.
For an $SFS$ type structure,
the proximity effect can lead to a modification of the LDOS that depends 
solely on the phase difference
of the superconductors\cite{bergeret}.
It was shown  by 
using a two-band model for ferromagnetism, 
that the LDOS for a $F/S$ bilayer still maintained the characteristic 
damped oscillations,
with however, an increased wavelength and sharper decline \cite{vedyayev}. 
Certain $F/S$ bilayers with
unconventional $d$-wave superconductors have also been investigated
for peculiar characteristics of the DOS \cite{melin,faraii,luck}.
Thus, while the DOS and LDOS
of $F/S$ bilayers have been investigated,
as have also to some extent $SFS$ junctions, 
work  analyzing the LDOS of junctions consisting of more than three layers
is still scarce. 
It is imperative therefore to have
a theory valid for these multilayers, since
as  potential device applications 
and experimental techniques become more advanced,
the creation of
complicated sequences of $\pi$-junctions will become increasingly realizable.

Much of the somewhat limited experimental work on the DOS in $F/S$ 
heterostructures
is  based on transport measurements. Direct evidence of the
oscillatory behavior of the superconducting correlations in the ferromagnet 
was found through tunneling spectroscopy measurements that showed inversions 
in the DOS
for a thin ferromagnet film \cite{kontos3}. Modifications to the DOS
in the superconductor also give insight 
in describing the proximity effect, 
by providing information on the influence of the
ferromagnet on superconducting correlations. It was found
that the DOS in the vicinity of the interface of the superconductor is
substantially modified from the bulk BCS result\cite{sillanpaa}.
Recently, local spectroscopy revealed the LDOS for
a few different ferromagnet thicknesses \cite{courtois3}:
the reported inversion of the DOS was attributed to
the $\pi$ state. 
Some other relevant experiments involving 
the ground state behavior of $\pi$-junctions have shown
that  both the $0$ or $\pi$ state can occur, depending on the width $d_F$
of the $F$ layer\cite{kontos}. The origin of the damped oscillatory behavior 
reflected
in DOS measurements and calculations discussed earlier 
is responsible, in some cases, for the 
damped oscillations in 
the critical current $I_C$
as a function of $F$ layer width, suggesting  a $0$ to $\pi$ 
transition\cite{kontos2}.
The signature in the characteristic 
$I_C$ curves also indicates a 
crossover from
the $0$ to $\pi$ phase in going from higher to lower 
temperatures\cite{ryazanov}.
The current phase relation was also measured\cite{frolov},
demonstrating a re-entrant $I_C$ with temperature variation.

Despite the existence of this experimental work, little attention has
been paid until very
recently to discussing, from a  thermodynamic
point of view,  the relative stability of the different   possible
configurations involving $0$ and $\pi$ junctions in $SFS\ldots$ 
heterostructures. Indeed, a recent preprint\cite{mon}
complains  of the complete absence of work on thermodynamic properties 
of such structures. We have alleviated this situation in very
recent work\cite{stab}, where we have found a way to correctly
evaluate the condensation free energy in these nanostructures as a function
of the relevant parameters. The work that we discuss here involves
discussing the LDOS for junction configurations
whose thermodynamic stability properties are\cite{stab} well understood. 

The quasiparticle amplitudes and energies 
that ultimately govern all observable behavior 
are very sensitive to the geometry
of the given $F/S$ system. For $F$ or $S$ layers 
a few nanometers thick, interference of the wavefunctions 
even over the atomic scale
can produce significant contributions to the LDOS. Often
the interaction potentials and geometry  
vary quite rapidly over the Fermi wavelength,
and averaging over the momentum space trajectories
eliminates useful information on the quasiparticle dynamics.
It is therefore preferred in the clean limit situation
considered here, to
implement a microscopic model which affords the most complete information regarding %kh
the local electronic structure.
This requires going beyond the standard
approximate equations, e.g., the
Eilenberger equations, which  are essentially transport differential equations 
that
describe superconductivity in the quasiclassical regime.
With the progressive strides made in local spectroscopy techniques, 
a fully microscopic theory that allows inclusion of the
details of the LDOS at the atomic scale
is necessary. Furthermore the ground state properties of $\pi$ junctions
can be analyzed properly only if the superconducting order parameter
is calculated self-consistently. The theory used here satisfies
these requirements.
  
In this paper we calculate  the LDOS for several 
possible 
multilayer structures. The existence of such structures
as either local or global minima of the free energy was
proved in Ref.~\cite{stab}.
Differing spectra are found depending on the exact 
nature
of the 
particular junction configuration considered. 
By using an established fully microscopic method \cite{hv3,hv1,hv2}, we
self-consistently calculate the pair potential, $\Delta({\bf r})$,
revealing several permissible types of
spatial arrangements of the order parameter.  
The quasiparticle wavefunctions and associated
energies are obtained from the microscopic Bogoliubov-de Gennes (BdG) equations.
A numerical algorithm is then used that permits the self-consistent calculation
of the pair potential.
We find differing signatures for the LDOS
that should be discernible in  
STM spectroscopy experiments.

\section{Methods} \label{meth}
In this section we outline  the Bogoliubov-de Gennes (BdG) equations \cite{bdg} 
that are appropriate 
for the investigations of $F/S$ multilayered junctions.
Since the BdG equations are 
inherently real-space based, they are
especially convenient for the 
investigation of the possible
spatial configurations  of the pair amplitude
and the spatially resolved LDOS.

We consider 3-D slab-like compounds 
translationally 
invariant
in  the $x-y$ plane, and with all spatial variations occurring in the $z$ direction.
The structures consists of alternating superconducting,  $S$, and ferromagnetic,
$F$, layers, each of width $d_S$
and $d_F$ respectively. We call $d$ the total thickness of the slab.
The corresponding coupled equations for the spin-up and spin-down quasiparticle
amplitudes ($u_{n}^{\downarrow},v_{n}^{\uparrow}$) are written,
%\begin{widetext} 
\begin{subequations}\label{bogo}
\begin{align}
 [-\frac{1}{2m}\frac{\partial^2}{\partial z^2} +\varepsilon_{\perp} 
-E_F(z) + U(z)- h_0(z)]u_n^{\uparrow}(z)+\Delta(z)v_n^{\downarrow}(z)&=\epsilon_n u_n^{\uparrow}(z)\\
-[-\frac{1}{2m}\frac{\partial^2}{\partial z^2} +\varepsilon_{\perp} 
-E_F(z) + U(z)+ h_0(z)]v_n^{\downarrow}(z)+\Delta(z)u_n^{\uparrow}(z)&=\epsilon_n v_n^{\downarrow}(z),
\end{align}
\end{subequations}
%\end{widetext}
where $\varepsilon_{\perp}$ is  the kinetic 
energy term corresponding to quasiparticles with momenta transverse to
the $z$ direction, $\epsilon_n$ are the 
energy eigenvalues,  $\Delta(z)$ is the pair potential, and $U(z)$
is the  scattering potential  at each $F/S$ interface, given by:
\begin{equation}
U(z)=\sum_{i=1}^{{(N_L-1)}/{2}}[U(i(d_S+d_F),z)+U((i(d_S+d_F)-d_F),z)],
\end{equation}
where we take $U(z_l,z)\equiv H \delta(z-z_l)$, $z_l$ is the location
of the interface, $H$ is the scattering parameter, and
$N_L$ represents
the total number of layers (superconducting plus 
magnetic). 
The  ferromagnetic exchange energy $h_0(z)$ 
takes the constant value $h_0$
in the $F$  layers, and zero elsewhere.
%The potential $U(z)$ represents the scattering at the $F/S$ interfaces.
Other relevant material parameters are taken into account through
the variable bandwidth
%Fermi energy,
$E_F(z)$. This is
taken to be $E_F(z)=E_{F S}$ in the
$S$ layers, while in the $F$ layers
one has $E_F(z)=E_{FM}$ 
so that in these regions the up and down
bandwidths are respectively $E_{F\uparrow}=E_{FM}+h_0$, and 
$E_{F\downarrow}=E_{FM}-h_0$. The dimensionless parameter
$I$, defined as $I\equiv h_0/E_{FM}$, conveniently characterizes
the magnets' strength. 
The ratio $\Lambda \equiv E_{FM}/E_{FS}
\equiv (k_{FM}/k_{FS})^2$
describes the mismatch between Fermi wavevectors on the $F$ and $S$
sides, assuming parabolic bands with $k_{FS}$ denoting
the Fermi wave vector in the $S$ regions. 
The dimensionless parameter $H_B\equiv mH/k_{FS}$ 
characterizes
the interfacial scattering strength. 
We consider here singlet pairing only and, consistent with this
choice, we assume
that the magnetic orientation in all magnetic
layers, when more than one is present, is the same. 

The  nontrivial 
spatial dependence of the pair potential must be calculated in a self
consistent manner by an appropriate sum over states:
%\begin{widetext}
\begin{equation}  
\label{del2} 
\Delta(z) = \frac{\pi g(z) N(0)}{k_{FS} d}\sum_{|\epsilon_n|\leq \omega_D}\int{d\varepsilon_\perp}
\left[u_n^\uparrow(z)v^\downarrow_n (z)+
u_n^\downarrow(z)v^\uparrow_n (z)\right]\tanh(\epsilon_n/2T),
\end{equation} 
%\end{widetext}
where  $N(0)$ is the density of states (DOS) per spin of the 
superconductor
in the normal state, $d$ is the total
system size in the $z$ direction, $T$ is the temperature, $\omega_D$ is the 
cutoff ``Debye'' energy of the pairing interaction,
and $g(z)$ is the effective  coupling, which we take to be a constant
$g$ within the  superconductor regions and zero elsewhere.
The quasiparticle amplitudes,
$u_{n}^{\downarrow}$ and $v_{n}^{\uparrow}$ involved in the sum 
are extracted from the solution to (\ref{bogo}) 
by using
symmetry arguments \cite{hv1}.

An appropriate choice of basis allows
Eqs.~(\ref{bogo})
to be 
transformed into a finite $2N \times 2N$ dimensional matrix eigenvalue 
problem in wave vector space:
\begin{equation}
\label{mbogo}
\begin{bmatrix} H^+&D \\
D & H^- \end{bmatrix}
\Psi_n
=
\epsilon_n
\,\Psi_n,
\end{equation}
where 
$\Psi_n^T =
(u^{\uparrow}_{n1},\ldots,u^{\uparrow}_{nN},v^{\downarrow}_{n1},
\ldots,v^{\downarrow}_{nN})$, are the expansion coefficients
associated with
the set of orthonormal basis vectors,
$u^{\uparrow}_n(z)=\sqrt{{2}/{d}}\sum_{q=1}^N u^{\uparrow}_{n q}\sin(k_q z)$, and
$v^{\downarrow}_n(z)=\sqrt{{2}/{d}}\sum_{q=1}^N v^{\downarrow}_{n q}
\sin(k_q z)$. The longitudinal momentum index $k_q$
is quantized via
$k_q = {q/\pi}{d}$, where
$q$ is a positive integer.
The label $n$ encompasses a sum over the index $q$
and the value of $\varepsilon_\perp$. 
The finite range of the pairing interaction $\omega_D$, implies
that $N$ is finite.
We have calculated 
the  matrix elements in Eq. (\ref{mbogo}) elsewhere \cite{stab}, 
and for brevity, the results are suppressed here.

We are mainly interested 
in 
the quantity of 
most 
experimental relevance: the local density of states (LDOS).
The LDOS is related to
the convolution of one-particle spectral functions for the
$S$ and $F$ regions, and is
defined as
$N(z,\varepsilon)= N_\uparrow(z,\varepsilon)+N_\downarrow(z,\varepsilon)$,
where the LDOS for
each  spin orientation is given by,
\begin{equation}\label{edos}
{N}_\sigma(z,\epsilon) 
=-\sum_{n}
\Bigl\lbrace[u^\sigma_n(z)]^2
 f'(\epsilon-\epsilon_n) 
+[v^\sigma_n(z)]^2
 f'(\epsilon+\epsilon_n)\Bigr\rbrace.  
\end{equation}
Here $\sigma=\uparrow,\downarrow$ and
$f'(\epsilon) = \partial f/\partial \epsilon$ is the derivative
of the Fermi function.

\section{Results} \label{res}

In this section we present results for the LDOS and for the  
pair amplitude  $F({\bf r})=\langle \hat{\psi}_{\downarrow}({\bf r})
\hat{\psi}_{\uparrow}({\bf r})\rangle$, where the $\hat{\psi}_\sigma$ are the
usual annihilation operators.
The results are given as a function  of the dimensionless 
parameters $I$, $H_B$ and $\Lambda$, as indicated. The geometrical parameters
will be given in dimensionless form in terms of the inverse of $k_{FS}$:
$D_S\equiv k_{FS} d_S$ and $D_F\equiv k_{FS} d_F$.  We take $D_F=10$ and
$d_S=\xi_0$, where $\xi_0$ is the BCS coherence length, with
$D_S=100$. The cutoff frequency is fixed to $\omega_D=0.04 E_{FS}$. From
these values and standard relations one infers $g$. The temperature
will be fixed to $T=0.01 T_c^0$, where $T_c^0$
is the bulk transition temperature of $S$.
All results will be
presented  in convenient dimensionless form as indicated.

Our  equations (including the self consistent
condition) are numerically solved using a previously 
described\cite{hv3,stab,hv1,hv2} iteration process.
The different possible self consistent states are found by starting
the iteration process with  initial guesses of different types. For
example, for an $SFS$ structure one can start either with a ``0'' or
a ``$\pi$'' initial guess. The resulting self consistent state reached
after iteration and convergence will be of the initial type, if such a
self consistent state exists at least
as  a metastable state (a local
minimum of the free energy) 
for the parameter values and geometry under
consideration: otherwise\cite{stab} the process converges to a different 
configuration. 

We consider three layer $SFS$ and seven layer $SFSFSFS$ structures. In the
first case, there is only one junction, and it can be in either the 0 or
the $\pi$ state. In the second case, there are three junctions, each
of which can in principle be in a 0 or $\pi$ state. Each overall junction
configuration can then be characterized by specifying the three values,
each zero or $\pi$, corresponding to the state of each junction.

\subsection{Structures} \label{struc}
\begin{figure}
\includegraphics[scale=.75,angle=0]{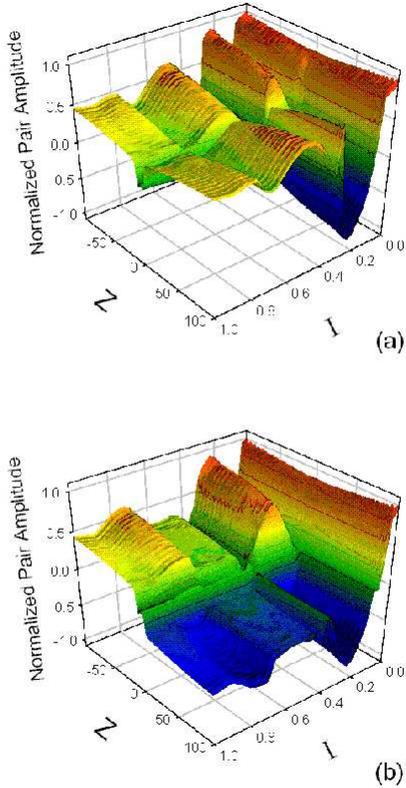}
\caption{\label{fig1} The pair amplitude $F(Z)$, normalized
to its bulk value $F_0$, for a three  
layer $SFS$
structure, plotted as a function of $Z\equiv k_{FS} z$ 
and of the dimensionless
magnetic polarization $I$, at $\Lambda=1$ and $H_B=0$.
The $Z=0$ point is at the center of the structure. 
Panel (a)
corresponds to self consistent results obtained (see text)
with an initial guess where the
junction is of the ``0'' type and panel (b) with
a ``$\pi$'' type. 
In the latter case, the solution found is  of the
$\pi$ type except at very small $I$ and in the neighborhood
of $I\approx 0.4$, but in the former case (panel (a)), a more
complicated behavior occurs, as discussed in the text.
}
\end{figure}

In this subsection 
we  present some results for the 
superconducting correlations  in $SFS$ and $SFSFSFS$ structures, as described
by the
pair amplitude, $F(z)$.
We do this to illustrate  the possible
self-consistent pair amplitude configurations that arise
as parameters such as the interface scattering 
strength $H_B$, the Fermi wavevector mismatch $\Lambda$, and the 
magnetic polarization strength $I$ vary.  This short discussion 
will establish what the different structures encountered
look like, some of their stability properties, and our notation.
The presentation is most clearly and efficiently illustrated
by means of 3-D plots.

\begin{figure}
\includegraphics[scale=.75,angle=0]{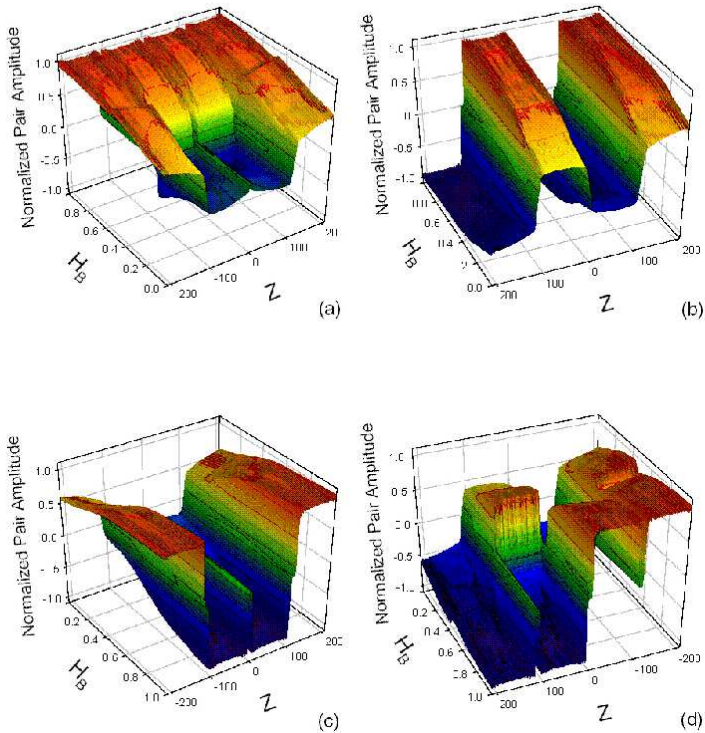}
\caption{\label{fig2} The normalized pair amplitude $F(Z)$  
for a  $SFSFSFS$ seven layer structure, plotted as
a function of $Z$ and of the barrier height parameter $H_B$.
Here $I=0.2$ and $\Lambda=1$.
Panels (a) (b) (c) and (d) correspond, respectively, 
to attempts to find solutions of the $000$, $\pi\pi\pi$, $\pi0\pi$ and
$0\pi0$ types (see text). The resulting structure, however, it is
not necessarily of the sought type, since some of
the structures are unstable in part of the $H_B$ range,
as explained in the text. For clarity, the direction of increasing
$H_B$ is not the same in every panel.}
\end{figure} 

To begin, we display, in 
Fig.~\ref{fig1},  the pair amplitude (normalized to $F_0$,
its bulk superconductor value) 
for a three layer $SFS$
system, 
as a function of position and
of the magnetic exchange parameter $I$ in their entire ranges, without
a barrier or mismatch ($\Lambda=1$ and $H_B=0$). 
The top panel shows the results of attempting to find a solution of the
0 type by starting the iteration process
with an assumed piecewise constant of that form. Whether this
attempt succeeds or not is reflected, of course, in the
self consistent results plotted. 
Careful examination of the two panels
reveals a rather intricate situation:  one can see
in the bottom panel that a solution of the $\pi$ type
exists nearly in the entire  $I$ range, the 
exception being at very small
$I$, where the effect of magnetism becomes very
weak and, as one would expect, only the $0$ state solution is found.  This
requires  small values of $I$,  $I \lesssim 0.1$ however. One can see that
near this small value, as the 0 state transitions to a $\pi$ configuration,
the amplitude of $F(z)$ is very small throughout the sample, which should
be reflected by a sharp dip in the transition temperature. 
A brief instability of the $\pi$ state also occurs in a small interval around 
$I\approx0.4$, where the self consistent solution is of the 0 type. This
instability is reflected in Fig.~7 of Ref.~\cite{stab}.
On the other
hand (panel (a)), one can see that there is a  region of $I$ 
values, $0.12 \lesssim I \lesssim 0.3$, where a 0 state does not exist.
The attempt to find it fails, and the solution converges, after a very large
number of iterations, to the same self consistent $\pi$ state that is found,
in the same $I$ range, in panel (b).

\begin{figure}
\includegraphics[scale=.75,angle=0]{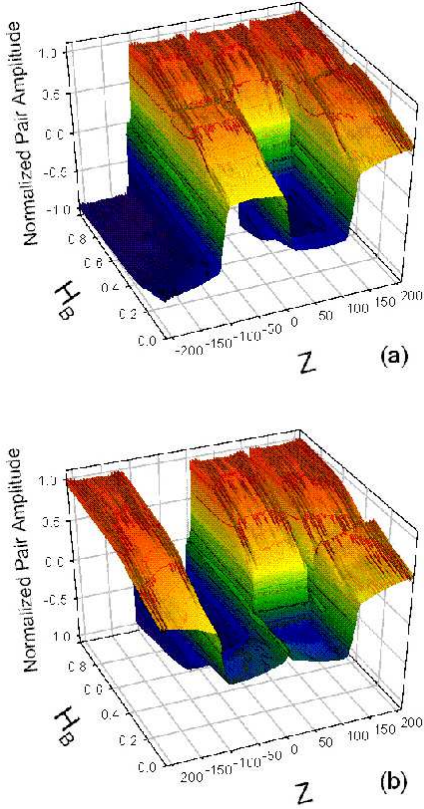}
\caption{\label{fig2a} The normalized pair amplitude $F(Z)$  
for a  $SFSFSFS$ seven layer structure, as in the previous Figure. 
The top and bottom panels correspond to initial attempts to find the 
asymmetric $\pi00$ and $\pi\pi0$
states respectively.
Clearly such attempts only succeed when the barrier is high and 
the proximity effects weak (see text).}
\end{figure}

\begin{figure}
\includegraphics[scale=.75,angle=0]{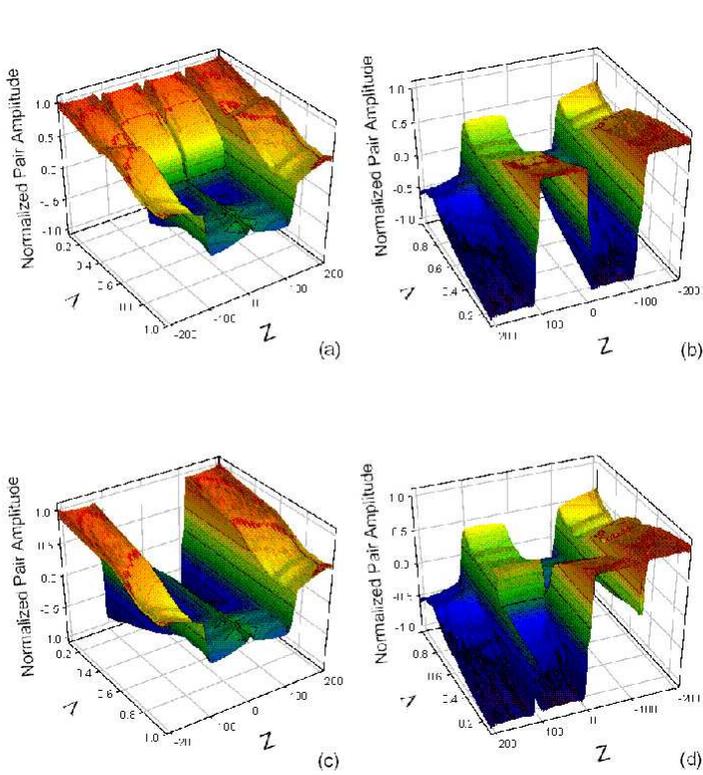}
\caption{\label{fig3} The normalized pair amplitude $F(Z)$  
for a  $SFSFSFS$ seven layer structure, plotted as
in  Fig.~\ref{fig2} and with the same panel arrangements,
but as a function of the mismatch parameter $\Lambda$ at zero
barrier. Note that the direction of increasing $\Lambda$ is, for
clarity, not the same in all the panels.
 }
\end{figure}

\begin{figure}
\includegraphics[scale=.75,angle=0]{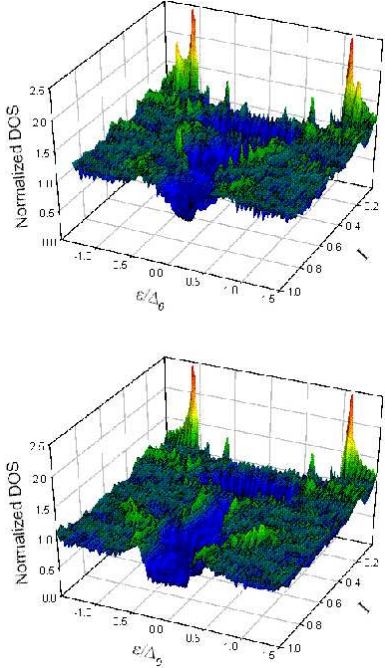}
\caption{\label{fig3a}  The normalized DOS (see
text) for a  $SFS$ trilayer, plotted as
a function of the energy (in units of the bulk zero temperature gap),
and of the exchange energy parameter $I\equiv h_0/E_{FS}$.   
The panel arrangement is the same as in 
Fig.~\ref{fig1}: therefore the plots
correspond to 0 and $\pi$ self consistent
states as in that figure.} 
\end{figure}

\begin{figure}
\includegraphics[scale=.75,angle=0]{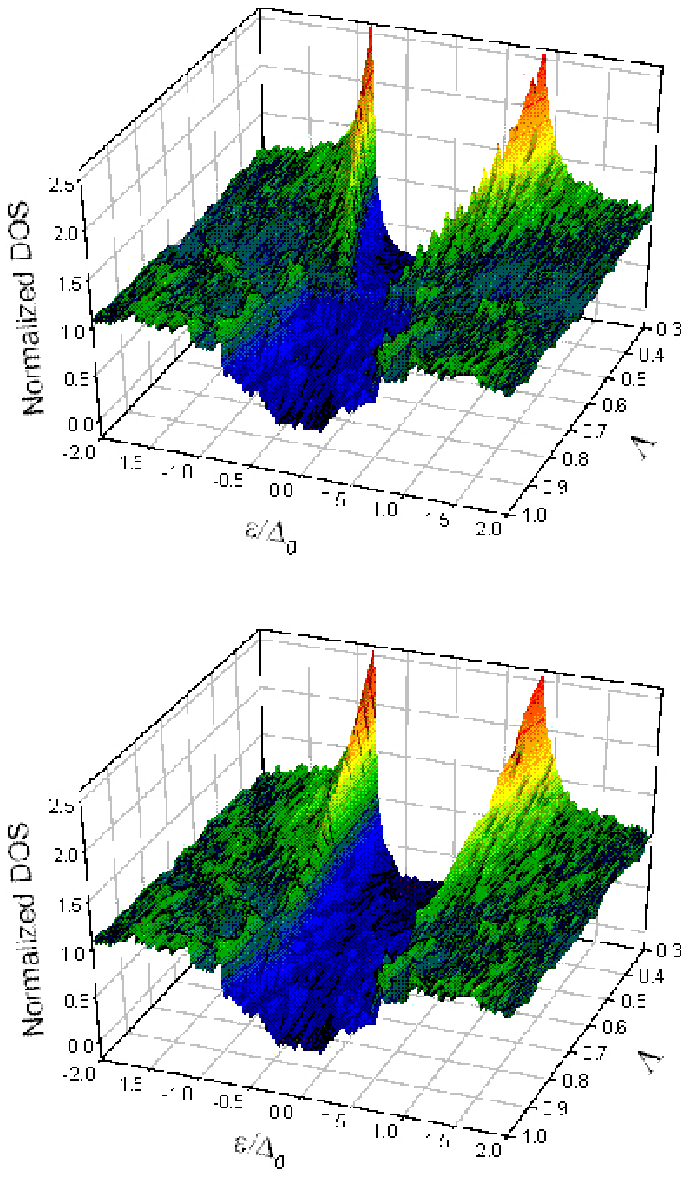}
\caption{\label{fig4}  The normalized DOS 
for a  $SFS$ trilayer, plotted as
a function of the energy,
and of the mismatch parameter $\Lambda$, at $I=0.2$ and $H_B=0$.  
The panel arrangement again corresponds to that in 
Fig.~\ref{fig1}.  
In the bottom panel, the results plotted are for a self consistent $\pi$
structure, while in the top panel they are for a $0$ structure
for $\Lambda <0.64$ and a $\pi$ structure at smaller mismatch, when
the 0 structure is not stable.}
\end{figure}

We now  illustrate the behavior of
$F(z)$ in  seven layer $SFSFSFS$ structures, which 
can be viewed as consisting  of 
three adjacent $SFS$ junctions. In the notation
described above we
denote as ``$000$" the junction
configuration when adjacent $S$ layers always have the same 
sign of $\Delta(z)$,  and as ``$\pi\pi\pi$" when
 successive
$S$ layers alternate in sign.
There are up to a trivial reversal, two more symmetric states: one in which 
$\Delta(z)$
has the same sign in the first
two $S$ layers, while in the last two it has the opposite sign,
(this is labeled as the ``$0\pi0$" configuration), and the other 
corresponding to
the two outer $S$ layers having the same sign for $\Delta(z)$, 
opposite to that of the two inner $S$ layers: these are 
referred to as ``$\pi0\pi$" junction configurations in our notation.
We will mainly focus our study on these symmetric configurations. 
For the first example
we consider the effect of the barrier thickness, as determined by the
parameter $H_B$.
This is done in Fig.~\ref{fig2}.  In this figure we
have taken $I=0.2$ and $\Lambda=1$ (no mismatch). The geometrical
parameters and temperature are the same as in Fig.~\ref{fig1}.
The panels (a), (b), (c)
and (d) correspond (in this order) to attempts to find solutions
of the 000, $\pi\pi\pi$, $\pi 0 \pi$, and $0\pi 0$ type. One can see that
in all cases each attempt succeeds when $H_B$ is sufficiently large, (the
direction of increasing $H_B$ has been reversed in two of the panels so
that this feature is not concealed by the perspective).
This makes  sense, as at larger
barrier heights the layers become more isolated from each other and the 
proximity effects must  eventually become insignificant.  The relative
phase of $\Delta(z)$ at each $S$ layer should then become
progressively irrelevant. For the same reason, 
there is a clear increasing trend with
increasing $H_B$ for the absolute values of $F(Z)$ in the middle
of the S layers. At smaller values
of $H_B$, a more careful examination of the panels 
reveals a rather intricate situation: two of the 
junction configurations, $\pi\pi\pi$ and $\pi 0\pi$ (panels (b) and (c)
respectively), exist in the entire range of $H_B$, while 
attempts to find a 000 structure
(panel (a)) result in a  $\pi 0\pi$ configuration for
small barrier heights $H_B \lesssim 0.48$, and similarly, attempts to find
a $0\pi0$ structure  result in a self consistent $\pi\pi\pi$ configuration
for $H_B \lesssim 0.44$. This is in perfect agreement with the
stability analysis of Ref.~\cite{stab}. Both of these ``transitions''  
consist simply of the flipping of the two outer junctions, with the inner
one remaining in the same configuration.

For completeness,
we show in Fig.~\ref{fig2a}, results for the two possible
asymmetric states that can exist ($\pi00$ and $\pi\pi0$). 
The parameter values
are identical to the symmetric case exhibited in the previous figure.
The top panel (a) corresponds to the self-consistent profile
when an assumed form for the pair potential corresponds to 
a $\pi 0 0$ junction, while the bottom panel is for
a $\pi \pi 0$ initial guess.
It is evident that, for reasons specified above, at large enough
values of the scattering parameter $H_B$ both asymmetric states 
are stable self-consistent solutions. We find that the $\pi00$ state is
stable for $H_B \gtrsim 0.52$, while the $\pi\pi0$ junction is more robust,
retaining its asymmetric configuration over a larger range of $H_B$ values,
$H_B\gtrsim 0.35$. However, for weaker barriers, where the proximity
effects are important, these asymmetric states are not stable, becoming
$\pi\pi\pi$ and $\pi0\pi$ respectively.

\begin{figure}
\includegraphics[scale=.75,angle=0]{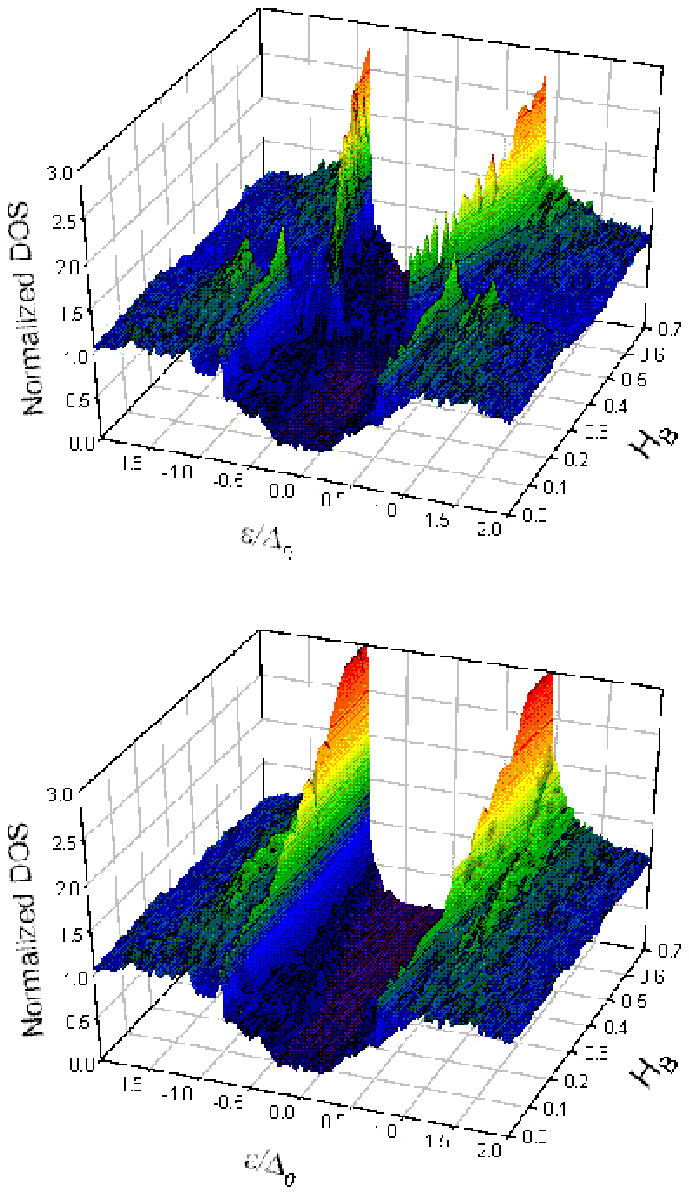}
\caption{\label{fig5}  The normalized DOS  for a  $SFS$ trilayer, plotted 
and displayed as
in Fig.~\ref{fig4}, and under the
same conditions, but as a function of barrier height rather than mismatch.
The results shown in the bottom panel correspond to a $\pi$ 
structure, while those the top panel they are for a $0$ structure
for $H_B>0.29$ and a $\pi$ structure at smaller barrier height values, when
the 0 structure is not stable (see text).}
\end{figure}

In Fig.~\ref{fig3} we consider the influence of the Fermi
wavevector mismatch,
characterized by the parameter $\Lambda\equiv E_{FM}/E_{FS}$. This figure is
completely analogous to Fig.~\ref{fig2}, except for the substitution of
$\Lambda$ for $H_B$, which is set
to zero (no barrier). We see   that
for sufficiently large mismatch (small $\Lambda$), all four junction
configurations exist. This is because increasing the mismatch does, in
effect, isolate the superconducting regions from each other just as efficiently
as increasing the barrier does. This can be roughly understood 
qualitatively from simple
quantum mechanical considerations. For smaller mismatch, only the 
two solutions of the $\pi\pi\pi$ and $\pi0\pi$ type exist: again
000 turns into $\pi 0\pi$ (for $\Lambda \gtrsim 0.48$) while $0\pi 0$ turns
into $\pi\pi\pi$ at $\Lambda \gtrsim 0.57$, these results being in agreement \cite{stab}
with those of detailed stability studies.

\subsection{DOS} \label{dos}

\begin{figure}
\includegraphics[scale=.5,angle=0]{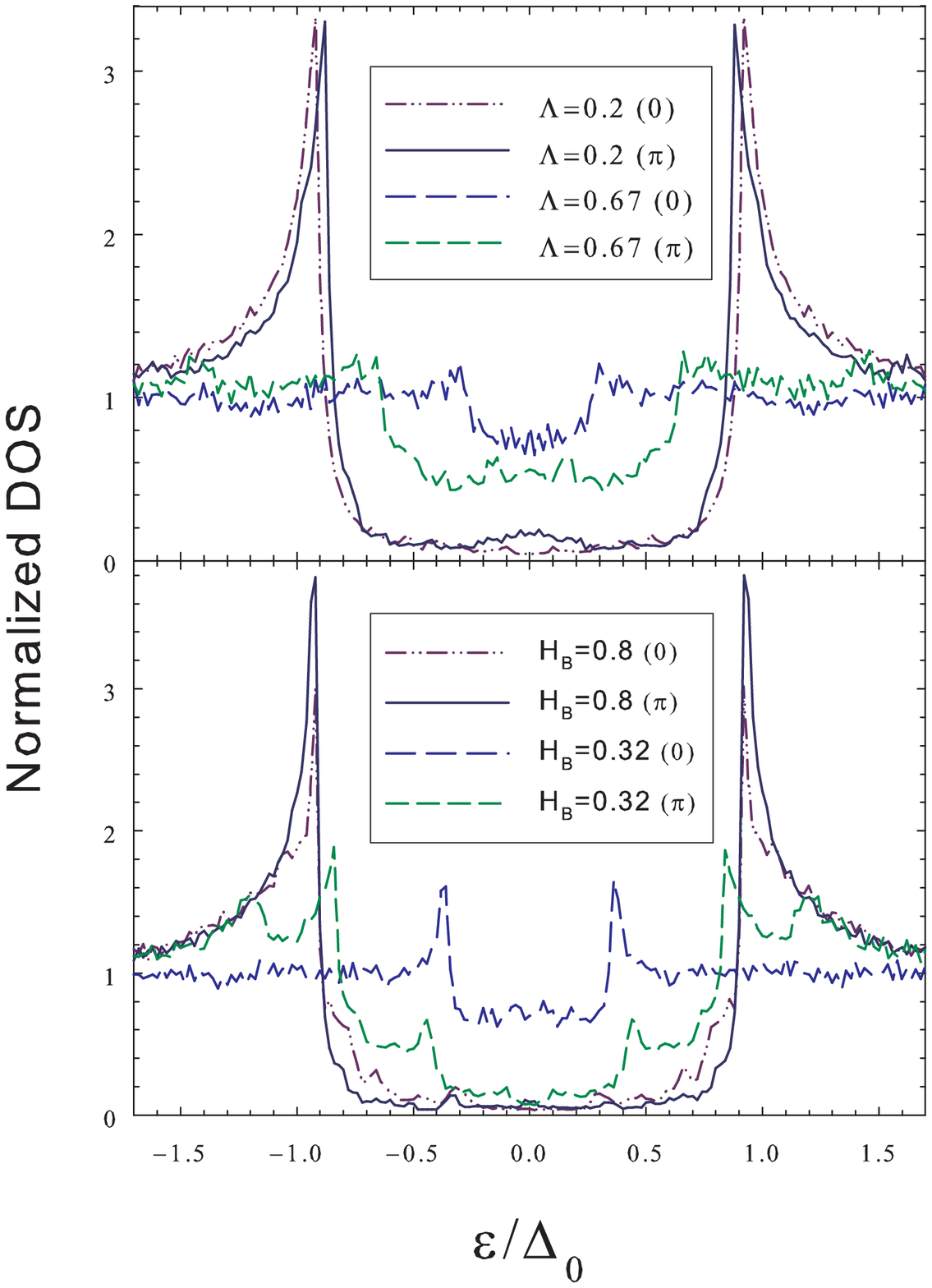}
\caption{\label{fig6}  The normalized DOS 
for a  $SFS$ trilayer, plotted as
a function of the energy.
Results for both 0 and $\pi$ self consistent
states are given, as indicated. These are slices from Figs.~\ref{fig4}
and \ref{fig5}.
In the top panel, the DOS is shown at $H_B=0$ 
for two different values of the mismatch
parameter, $\Lambda=0.2$, and 0.67,
the latter being a case for which the 0 state is nearly unstable
(see text).
The bottom panel shows the DOS profile in the absence of mismatch 
($\Lambda=1$), but with
the interface scattering parameter $H_B$ taking on the two values shown,
chosen on similar criteria as the $\Lambda$ values (see text).
 }
\end{figure}

We now present some of our results for the LDOS, which is the main
focus of our paper.
Although the quantity $F(z)$ gives useful information regarding the superconducting correlations
in the structure, it is the LDOS
which is 
experimentally measured. 
The results given are in all cases for the quantity ${N}_\sigma(z,\epsilon)$
defined in Eq.~(\ref{edos}), 
summed over $\sigma$, and 
integrated over 
a distance of one coherence length (or, equivalently, the
thickness of one superconducting layer) from the edge of the sample.
We  normalize our results to the
corresponding value for the normal state DOS of the superconducting material,
and the energies to the bulk value of the gap at zero
temperature, $\Delta_0$.
The results shown are for the three and seven layer geometries discussed
at the beginning of this section.
Material   parameters not otherwise mentioned
are set to the ``default'' values of $H_B=0$, $\Lambda=1$ and $I=0.2$

Figure \ref{fig3a} characterizes the sensitivity of the LDOS,
in a $SFS$ junction, to
variations in the dimensionless exchange field parameter, $I$.
The panel arrangement  is as in Fig.~\ref{fig1}.
The top panel, therefore, shows the resultant DOS when the structure is 
predominately
in the $0$ state configuration, the exception being a small range of I
as discussed in conjunction with Fig.~\ref{fig1}, where the 0 state
is unstable. 
The bottom panel shows the results for the
$\pi$ junction case, except at very small $I$.
The limiting case of an $SNS$ junction ($I=0$) displays 
(as can be seen in
either panel)
subgap Andreev bound states that 
quickly decay as $I$ increases. 
Upon varying 
$I$, the number of subgap states   oscillates 
as a function of $I$, 
in a manner that is dependent on whether the stable state is $0$ or $\pi$.

In Fig.~\ref{fig4} we show some
other  results for the same three layer system. The figure shows
the normalized local
density of states, as specified above, as a function of  energy and
mismatch $\Lambda$,
at fixed $I=0.2$. The panel arrangement is again as in Fig.~\ref{fig1}:
the bottom
panel represents the results for a $\pi$ configuration, while the top panel
shows the results for the 0 configuration
at larger mismatch, $ \Lambda<0.64$, while
for smaller mismatch, where the 0 configuration does not\cite{stab}
exist, it repeats the
results of the $\pi$ configuration. The change in the results reflecting
the onset of the 0 configuration instability at
$\Lambda=0.64$ can be discerned in panel (a).
At larger mismatch (smaller $\Lambda$, the back portion
of the figures), where both states can coexist, one can
clearly see in the figure the difference between the two configurations.
This is helped because in general 
when there is more mismatch the
proximity effect weakens, the gap opens and larger peaks form which 
progressively become more BCS-like as the mismatch increases ($\Lambda$
decreases) to $\Lambda=0.2$. The way this occurs, too, is not the same
for the 0 and the $\pi$ configurations. 

\begin{figure}
\includegraphics[scale=.75,angle=0]{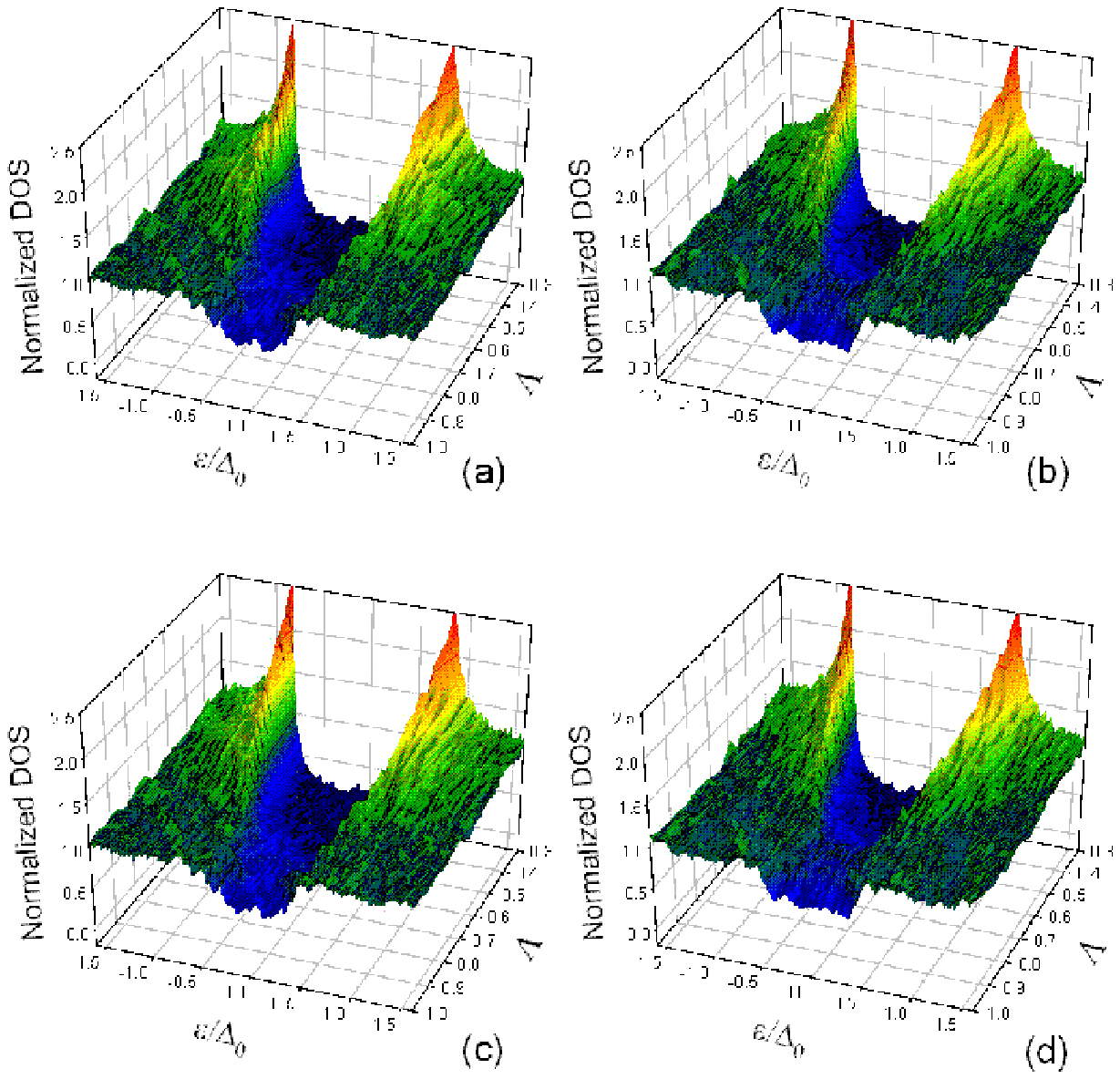}
\caption{\label{fig7}  The normalized DOS  
for the  $SFSFSFS$ structure of Figs.~\ref{fig2} and \ref{fig3}, 
with the same panel arrangements as in those figures, plotted as
a function of the energy
and of the mismatch parameter $\Lambda$. 
}
\end{figure} 

In the next figure, Fig.~\ref{fig5}, we repeat the previous study but in
terms of the barrier height, rather than the mismatch. Again, we plot in 
each panel the self consistent results obtained with initial guesses of the 0 
and $\pi$ types, which means that all regions of the (b) panel correspond
to self consistent $\pi$ states whereas those in the top panel correspond
to 0 configurations, except for $H_B<0.29$ where  no
self consistent 0 state exists\cite{stab}. 
We  then plot instead results 
for the  same $\pi$ state as in the bottom panel, which is reached by
iteration of either an initial 0 or $\pi$ guess. Again, the point where
the transition occurs can be seen as a clear discontinuity in panel (a).
The difference between the DOS for the
two states where  they coexist is in this case even more obvious than
in the previous one, with the BCS like peaks being considerably higher
for the $\pi$ configuration.

It is also illustrative to display
more clearly some of the features of the results
by isolating some selected slices of the 3-D
plots as 2-D line drawings. This we do in Fig.~\ref{fig6}.
In the top panel of this figure we present results for an $SFS$ 
trilayer, for
two contrasting values of the mismatch parameter $\Lambda$. 
Results labeled as ``0"  and  ``$\pi$" are for the
case where the self consistent states plotted are of these types.
The $0$ and $\pi$ state curves corresponding to $\Lambda=0.67$, where
(as shown above and in Ref.~\cite{stab}) the 0 state is barely metastable,
have clearly distinct signatures, with a smaller  gap opening for the $0$ 
state, and consequently
more subgap quasiparticle states.
Consistent with this, 
when there is little
mismatch one finds that the pair amplitude is relatively large
in the $F$ layer. 
In agreement  with what was seen in Fig.~\ref{fig4},
this progression takes a different form for the 0
and $\pi$ states, and this
is reflected in this panel, as can see by careful comparison of the curves. 
In the bottom panel we demonstrate the effect
of the barrier height. 
The results are displayed as in the top panel, but with the dimensionless
height $H_B$ taking the place of the mismatch parameter. This figure 
should be viewed in conjunction with 
Fig.~\ref{fig5}. One of the values
of $H_B$  chosen ($H_B=0.32$) is again such that the
0 state is\cite{stab} barely metastable, while for the other value 
the 0 and $\pi$ states  have similar condensation energies.
For
$H_B$ close to 0.32 there is a marked contrast between the two
plots, with the gap clearly opening  wider and containing more 
structure for the $\pi$ state. 
At larger $H_B$ the gap becomes larger in both cases, with
the BCS-like peaks increasing in  height.
Thus, as the barrier becomes larger, 
one is dealing with nearly independent superconducting slabs
and the plots become more similar. The largest difference therefore occurs, 
as for the mismatch, at the intermediate values more likely to be found
experimentally.

\begin{figure}
\includegraphics[scale=.75,angle=0]{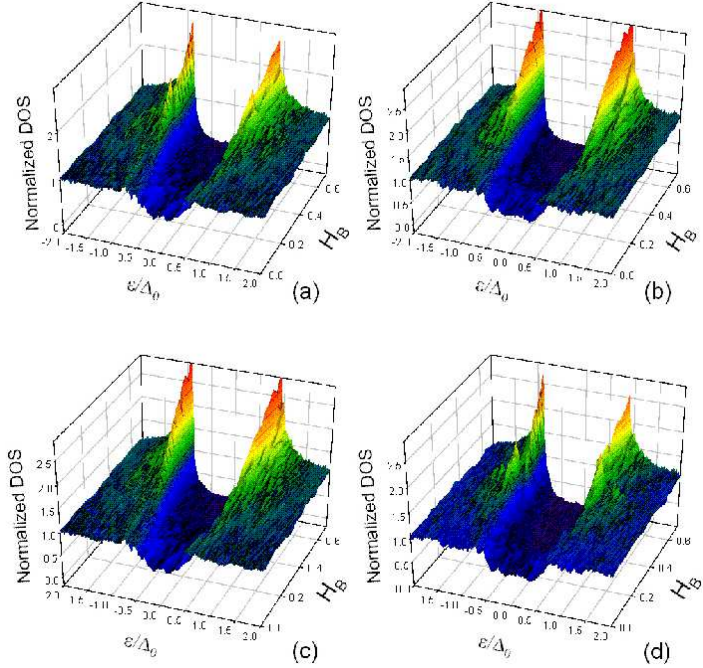}
\caption{\label{fig8}  The normalized DOS 
for the  $SFSFSFS$ structure of Figs.~\ref{fig2}, \ref{fig3}
and \ref{fig7}, plotted as
in Fig.~\ref{fig7} except as a function of barrier height $H_B$ instead
of $\Lambda$. 
}
\end{figure} 
Now we turn to the DOS results for the seven layer structure. In Fig.~\ref{fig7}
we present results for the normalized local DOS, defined as in the three
layer case, as a function of energy and mismatch parameter. The
panel arrangement in this figure corresponds to that in 
Fig.~\ref{fig3}, 
with however a lower bound of $\Lambda=0.3$ in the mismatch parameter, in order
to better
display the more intricate LDOS structure. The results shown are  
for the 
self consistent state reached by starting from each of the four symmetric 
junction configurations.  For larger mismatch, which corresponds
to the back portion of the plots,
all results are therefore
similar,
as the LDOS reflects the independent nature
of the isolated slabs (the phase is irrelevant in this limit).
For small mismatch (the ranges are indicated in the discussion
of Fig.~\ref{fig3}) the results of panel (a) coincide with those of panel (c)
and those of panel (d) with  those of (b), as can be seen
by comparing the front portions of these
panel pairs. The transitions
can be noted in the plots. Examination of the results shows
clear differences between the four configurations.
As in the three layer case, the peaks are
more prominent at larger mismatch, and the ``gap'' is wider.  
At small
mismatch, where only two
configurations are stable, the gaps fill up and
the peaks become very small for both cases, namely
$\pi\pi\pi$ (panel (b)  or large $\Lambda$
region of panel (d)), and $\pi 0\pi$ (panel (c) or front region
of panel (a)).

A similar situation occurs as a function of the barrier parameter $H_B$
as shown in Fig.~\ref{fig8}, which corresponds to the same conditions
as Fig.~\ref{fig2}, with the same conventions and panel arrangements.
In this case, however, the BCS peaks are (at larger barrier heights) more
prominent for the $\pi\pi\pi$ and $\pi 0\pi$ configurations since
the $S$ layers are not yet completely decoupled at the largest
$H_B$ shown, which (again for clarity) is smaller than that
used in Fig.~\ref{fig2}. 
The peaks weaken and the gap
narrows and fills up at small barrier heights, as the proximity
effect strengthens. The
prominence or weakening of the peaks is associated\cite{stab} with an
increase or decrease of the condensation energy of the corresponding
configuration, as one would intuitively expect. For small barrier heights
we note also that the two states that exist in this case ($\pi\pi\pi$ and
$\pi0\pi$) have again clearly 
different signatures, as can be seen by comparing the front (small $H_B$)
portions of diagrams (a) and (c) with those of (b) and (d). A similar
situation occurs at small mismatch (larger $\Lambda$), in Fig.~\ref{fig7}.

\section{Conclusions}\label{conclusion}

We have seen in this paper that the different junction configurations
that compete in stability for different ranges of the relevant parameters,
yield rather characteristic  
signatures in the local DOS. Thus,
our results illustrate that the complex and nontrivial
behavior of the pair amplitude in clean layered $F/S$ nanostructures
is reflected in the LDOS.
The numerical method used here is based upon
self consistent solution of the BdG equations.
One advantage of this method is that it properly takes into account 
the atomic-scale and 
quantum interference effects
that are likely to be important in these systems.
For the $SFS$ and $SFSFSFS$ structures investigated here, 
one can have up 
to two and six independent
locally stable configurations, respectively;  as 
the number of layers increases, so too does
the number of possible configurations.
Quantitatively, we have shown that 
the larger the scattering strength at the interfaces, or the mismatch
between Fermi wavevectors of the $F$ and $S$ regions,
the greater the number of stable states, since
the proximity effects weaken. The various pair amplitude profiles
however, exhibited distinct spatial forms in each case.
We confirmed\cite{hv3,stab,sc} 
that it is essential to calculate the pair potential 
self consistently,
since\cite{stab} as the results shown
here 
demonstrate, multilayer configurations can transition
from one state to another abruptly, as one varies any one of
the physical parameters: the magnetic exchange energy $I$, 
interface scattering strength $H_B$, 
and the Fermi wavevector mismatch $\Lambda$. 

The diverse 
behavior observed in the pair amplitude is 
found to be reflected in the calculated energy spectra.
When several self-consistent pair amplitude configurations coexist,
each representing a local or global minimum
of the free energy,  they yield specific signatures 
in the local DOS, which are susceptible to
measurement in local tunneling spectroscopy experiments.
For a trilayer $SFS$ junction, variations in the exchange
field parameter $I$ lead to 
oscillations in
the number of subgap states, and there are similar changes in the
seven layer spectrum.
We also investigated the LDOS profiles
by separately varying 
$H_B$ and $\Lambda$.
In the limiting cases of large $H_B$ or small $\Lambda$ 
the LDOS takes its typical BCS-like form, with few
subgap quasiparticle states. 
The LDOS in general however, has characteristic peaks and dips that 
depend on the symmetry of the ground state, and on
$H_B$, $I$ or $\Lambda$. 

It would be of obvious interest to examine whether or not the 
configuration changes
that occur as the material parameters are varied can happen also
as a function of temperature. This question is now under investigation.  

% Bibliographic references with the natbib package:
% Parenthetical: \citep{Bai92} produces (Bailyn 1992).
% Textual: \citet{Bai95} produces Bailyn et al. (1995).
% An affix and part of a reference:
%   \citep[e.g.][Ch. 2]{Bar76}
%   produces (e.g. Barnes et al. 1976, Ch. 2).


\begin{thebibliography}{5}

\bibitem{courtois} A.K. Gupta, L. Cretinon, N. Moussy, B. Pannetier, and H. Courtois,
Phys. Rev. B {\bf 69}, 104514 (2004)
\bibitem{courtois2} L. Cretinon, A. Gupta, B. Pannetier and H. Courtois,
Physica C {\bf 404}, 103 (2004).
\bibitem{and} A.F. Andreev, Zh. Eksp. Teor. Fiz. {\bf 46}, 1823 (1964)
[Sov. Phys. JETP {\bf 19} 1228 (1964)].
\bibitem{ff} P. Fulde and A. Ferrell, Phys. Rev.  {\bf 135}, A550 (1964).
\bibitem{lo} A. Larkin and Y. Ovchinnikov, 
Sov. Phys. JETP {\bf 20}, 762 (1965). 
\bibitem{bulaevskii} L.N. Bulaevskii, V.V. Kuzii, and A.A. Sobyanin, Pis'ma Zh. Eksp. Teor. Fiz. {\bf 25},
314 (1977) [JETP Lett. {\bf 25}, 290 (1977)].
\bibitem{hv3} K. Halterman and O.T. Valls, Phys. Rev. B {\bf 69}, 014517 (2004).
\bibitem{izy} Y.A. Izyumov, Y.N. Proshin, and M.G. Khusainov, Phys. Usp.
{\bf 45}, 109 (2002).
\bibitem{stab} K. Halterman and O.T. Valls, Phys. Rev. B {\bf 70}, 104516 (2004).
\bibitem{buzdin} A. Buzdin, Phys. Rev. B {\bf 62}, 11377 (2000).
\bibitem{zareyan} M. Zareyan, W. Belzig, and Y.V. Nazarov {\bf 86}, 308 (2001).
\bibitem{gap1} K. Halterman and O.T. Valls, Physica C {\bf 397}, 151 (2003).
\bibitem{gap2} D. Huertas-Hernando and Y.V. Nazarov, cond-mat/0404622 (unpublished).
\bibitem{koltai} J. Koltai, J. Cserti, and C.J. Lambert, Phys. Rev. B {\bf 69}, 092506 (2004).
\bibitem{vecino} E. Vecino, A. Mart\'{i}n-Rodero, and A.L. Yeyati Phys. Rev. B {\bf 64}, 184502 (2001).
\bibitem{buzdin2} I. Baladi\'{e} and A. Buzdin Phys. Rev. B {\bf 64}, 224514 (2001).
\bibitem{bergeret} F.S. Bergeret, A.F. Volkov, and K.B. Efetov, Phys. Rev. B {\bf 65}, 134505 (2002).
\bibitem{vedyayev} A. Vedyayev, A. Buzdin, D. Gusakova, and O. Kotelnikova cond-mat/0401037 (unpublished).
\bibitem{melin} N. Stefanakis, R, Melin, Journ. of Phys.-Cond. Mat. {\bf 15}, 3401 (2003).
\bibitem{faraii}  Z. Faraii and M. Zareyan, Phys. Rev. B {\bf 69}, 014508 (2004).
\bibitem{luck}  T. Luck T and U. Eckern,  Journ. of Phys.-Cond. Mat. {\bf 16}, 2071 (2004).
\bibitem{kontos3} T. Kontos, M. Aprili, J. Lesuer, and X. Grison, Phys. Rev. Lett. {\bf 86}, 304 (2001).
\bibitem{sillanpaa} M.A. Sillanp\"{a}\"{a}, T.T. Heikkila, R.K. Lindell, and P.J. Hakonen, 
Europhys. Lett. {\bf 56}, 590 (2001).
\bibitem{courtois3} L. Cretinon, A.K. Gupta, B. Pannetier, H. Courtois, H. Sellier, and F. Lefloch,
Physica C {\bf 404}, 110 (2004).
\bibitem{kontos} W. Guichard, M. Aprili, O. Bourgeois, T. Kontos, J. Lesueur, and P. Gandit,
Phys. Rev. Lett. {\bf 90}, 167001 (2003).
\bibitem{kontos2} T. Kontos, M. Aprili, J. Lesueur, F. Genet, B. Stephanidis, and R. Boursier,
Phys. Rev. Lett. {\bf 89}, 137007 (2002).
\bibitem{ryazanov} V.V. Ryazanov, V.A. Oboznov, A.Y. Rusanov, A.V. Veretennikov, A.A. Golubov,
and J. Aarts, Phys. Rev. Lett. {\bf 86}, 2427 (2001).
\bibitem{frolov} S.M. Frolov, D.J. Van Harlingen, V.A. Oboznov, V.V. Bolginov, and V.V. Ryazanov,
 cond-mat/0402434 (2004).
\bibitem{mon} J. Cayssol and G. Montambeaux, cond-mat/0404215 (2004).
\bibitem{hv1} K. Halterman and O.T. Valls, Phys. Rev. B {\bf 65}, 014509 (2002).
\bibitem{hv2} K. Halterman and O.T. Valls, Phys. Rev. B {\bf 66}, 224516 (2002).
\bibitem{bdg} P.G. de Gennes, {\it Superconductivity of Metals and Alloys} 
(Addison-Wesley, Reading, MA, 1989).
\bibitem{sc} K. Halterman and J.M Elson, J. Phys.: Condens. Matter {\bf 15}, 5837 (2003).
\end{thebibliography}
\end{document}